\documentclass[prb,a4paper,twocolumn,showpacs,floatfix,superscriptaddress,amsmath,amsfonts,amssymb,preprintnumbers,citeautoscript]{revtex4}
\pdfoutput=1
\usepackage[T1]{fontenc}\usepackage[latin1]{inputenc}
\usepackage{dcolumn,graphicx,color,booktabs,microtype,afterpage} \graphicspath{{Figures/}}
\usepackage[autoplay]{animate}
\usepackage[charter,greekuppercase=italicized]{mathdesign}\usepackage{sidecap}
\renewcommand{\figurename}{Fig.}
\renewcommand{\tablename}{Table}
\makeatletter\renewcommand{\fnum@figure}[1]{\figurename~\thefigure~(color online).}\makeatother
\makeatletter\renewcommand{\fnum@table}[1]{\tablename~\thetable.}\makeatother
\newcount\hh \newcount\mm
\hh=\time \divide\hh by 60
\mm=\hh \multiply\mm by 60 \mm=-\mm
\advance\mm by \time
\def\now{\number\hh:\ifnum\mm<10{}0\fi\number\mm}

\usepackage[colorlinks,plainpages=false,linkcolor=blue,urlcolor=blue,citecolor=blue,pdfpagemode=UseNone,pdfstartview=FitBH]{hyperref}

\newcommand{\half}{\frac{1}{\protect\raisebox{0.8pt}{\scriptsize 2}}}

\begin{document}

\makeatletter\renewcommand{\ps@plain}{%
\def\@evenhead{\hfill\itshape\rightmark}%
\def\@oddhead{\itshape\leftmark\hfill}%
\renewcommand{\@evenfoot}{\hfill\small{--~\thepage~--}\hfill}%
\renewcommand{\@oddfoot}{\hfill\small{--~\thepage~--}\hfill}%
}\makeatother\pagestyle{plain}

%\preprint{\textit{Preprint: \today, \now. For internal use only, do not distribute.}}%\linenumbers

\title{~\vspace*{-12pt}\\\hspace*{-1pt}\mbox{One-Dimensional Dispersive Magnon Excitation in the Frustrated Spin-2 Chain System Ca$_\text{3}$Co$_\text{2}$O$_\text{6}$}}

\author{Anil Jain}\email[\vspace{-5pt} Corresponding author: ]{ajain@barc.gov.in}
\affiliation{Max-Planck-Institut für Festkörperforschung, Heisenbergstraße 1, D-70569 Stuttgart, Germany}
\affiliation{Solid State Physics Division, Bhabha Atomic Research Centre, Mumbai 400085, India}

\author{P.~Y.~Portnichenko}
\affiliation{Institut für Festkörperphysik, TU Dresden, D-01069 Dresden, Germany}

\author{Hoyoung Jang}
\affiliation{Max-Planck-Institut für Festkörperforschung, Heisenbergstraße 1, D-70569 Stuttgart, Germany}
\affiliation{Stanford Synchrotron Radiation Lightsource, SLAC National Accelerator Laboratory, Menlo Park, California 94025, USA}

\author{G.~Jackeli}
\affiliation{Max-Planck-Institut für Festkörperforschung, Heisenbergstraße 1, D-70569 Stuttgart, Germany}

\author{G.~Friemel}
\affiliation{Max-Planck-Institut für Festkörperforschung, Heisenbergstraße 1, D-70569 Stuttgart, Germany}

\author{A.\,Ivanov}
\affiliation{Institut Laue-Langevin, 6 rue Jules Horowitz, F-38042 Grenoble Cedex 9, France}

\author{A.~Piovano}
\affiliation{Institut Laue-Langevin, 6 rue Jules Horowitz, F-38042 Grenoble Cedex 9, France}

\author{S.\,M.\,Yusuf}
\affiliation{Solid State Physics Division, Bhabha Atomic Research Centre, Mumbai 400085, India}

\author{B.~Keimer}
\affiliation{Max-Planck-Institut für Festkörperforschung, Heisenbergstraße 1, D-70569 Stuttgart, Germany}

\author{D.~S.~Inosov}\email[Corresponding author: \vspace{8pt}]{Dmytro.Inosov@tu-dresden.de}
\affiliation{Institut für Festkörperphysik, TU Dresden, D-01069 Dresden, Germany}
\affiliation{Max-Planck-Institut für Festkörperforschung, Heisenbergstraße 1, D-70569 Stuttgart, Germany}

\begin{abstract}%\parfillskip=0pt\relax%\linenumbers
\noindent Using inelastic neutron scattering, we have observed a quasi-one-dimensional dispersive magnetic excitation in the frustrated triangular-lattice spin-2 chain oxide Ca$_{3}$Co$_{2}$O$_{6}$. At the lowest temperature ($T=1.5$\,K), this magnon is characterized by a large zone-center spin gap of $\sim$\,27\,meV, which we attribute to the large single-ion anisotropy, and disperses along the chain direction with a bandwidth of $\sim$\,3.5\,meV. In the directions orthogonal to the chains, no measurable dispersion was found. With increasing temperature, the magnon dispersion shifts towards lower energies, yet persists up to at least $150$\,K, indicating that the ferromagnetic intrachain correlations survive up to 6 times higher temperatures than the long-range interchain antiferromagnetic order. The magnon dispersion can be well described within the predictions of linear spin-wave theory for a system of weakly coupled ferromagnetic chains with large single-ion anisotropy, enabling the direct quantitative determination of the magnetic exchange and anisotropy parameters.
\end{abstract}

\pacs{75.30.Ds 75.50.Ee 78.70.Nx 75.10.Pq\vspace{-0.7em}}

\maketitle\enlargethispage{3pt}

Frustrated antiferromagnets attract much theoretical and experimental attention because of their peculiar magnetic properties. The geometry of the underlying lattice or competing interactions in these systems may give rise to a macroscopic ground-state degeneracy and thus can prevent the onset of a long-range magnetic ordering (LRO) down to the absolute zero temperature, $T$\,=\,0\,K \cite{MoessnerRamirez06, Balents10}. Due to the large ground-state degeneracy, small perturbations, such as further-neighbor interaction, single-ion anisotropy, spin-lattice interactions, or magnetic field, can give rise to a variety of magnetically ordered states~\cite{LacroixMendels11}. Quasi-two-dimensional (2D) triangular-lattice antiferromagnets (TLAF) have been extensively studied as exemplar frustrated spin systems \cite{Mekata77, CollinsPetrenko97, NakatsujiNambu05, OlariuMendels06, TothLake12}. Their magnetic phase diagrams strongly depend on the exchange anisotropy. A typical 2D Heisenberg TLAF with nearest-neighbor interactions orders in the noncollinear ``$120^{\circ}$ structure'' even in the extreme quantum spin-$\half$ case \cite{BernuLhuillier92}. However, the 2D Ising TLAF is known to display no LRO down to $T$\,=\,0\,K~\cite{Wannier50}. So far, experimental realizations of the Ising TLAF are restricted to a few systems, where ferromagnetic (FM) \cite{Coldea177, Kimber104425, Lee702} or antiferromagnetic (AFM) \cite{YelonCox75} spin chains are arranged on a triangular lattice in the plane perpendicular to the chains. Among them, compounds with FM chains offer a rich playground to investigate a variety of exotic magnetic phases cooperatively induced by the dimensionality reduction, magnetic anisotropy, geometrical frustration, and magnetic field \cite{Coldea177, Lee702}.

As a model system of the Ising TLAF, Ca$_{3}$Co$_{2}$O$_{6}$ exhibits many intriguing properties, such as field-induced magnetization steps \cite{KageyamaYoshimura97, Hardy04, MaignanHardy04, FleckLees10, BakerLord11}, time-dependent magnetic order \cite{MoyoshiMotoya11, AgrestiniFleck11}, magnetodielectric coupling \cite{Bellido054430} and others \cite{ShimizuHoribe10, JainSingh06, HardyFlahaut07, Takubo073406, ChengZhou09, WeiHuang13}. It has a rhombohedral structure (space group $R\overline{3}c$) with a hexagonal arrangement of one-dimensional (1D) chains consisting of alternating face-sharing CoO$_{6}$ octahedra (OCT) and CoO$_{6}$ trigonal prisms (TP) with respectively low-spin ($S\,=\,0$) and high-spin ($S\,=\,2$) states of the Co$^{3+}$ ions \cite{Burnus245111, Takubo073406}. Besides, the large single-ion anisotropy of the Co$^{3+}_{\rm TP}$ ions leads to an Ising character of their spins, pointing along the \emph{c}-axis, while strong FM intrachain and weaker AFM interchain interactions combined with a triangular-lattice arrangement of spin chains give rise to a geometric frustration \cite{FresardLaschinger04}.

An effective 2D Ising model was initially proposed \cite{Kudasov027212, Yao212415} to explain the dc magnetization~\cite{KageyamaYoshimura97} and neutron-diffraction~\cite{Kageyama357, AaslandFjellwag97, PetrenkoWooldridge05} data. In this ``rigid chain'' model, every chain was replaced by a fictitious classical Ising spin. However, the recently observed incommensurability of the magnetic structure below the N\'eel temperature, $T_\text{N}\approx24$\,K, \cite{Bombardi100406, AgrestiniChapon08} and the time-dependent magnetic order-order transition from the incommensurate to the commensurate half-integer AFM structure below $\sim$\,10\,K~\cite{AgrestiniFleck11} have challenged the validity of this model. Consequently, a 3D lattice spin model was proposed \cite{Chapon172405, KamiyaBatista12}, in which the nearest- and next-nearest-neighbor interchain AFM interactions were assumed to compete with the dominant FM intrachain coupling.

\begin{figure*}[t]\vspace{-1.2em}
\includegraphics[width=\textwidth]{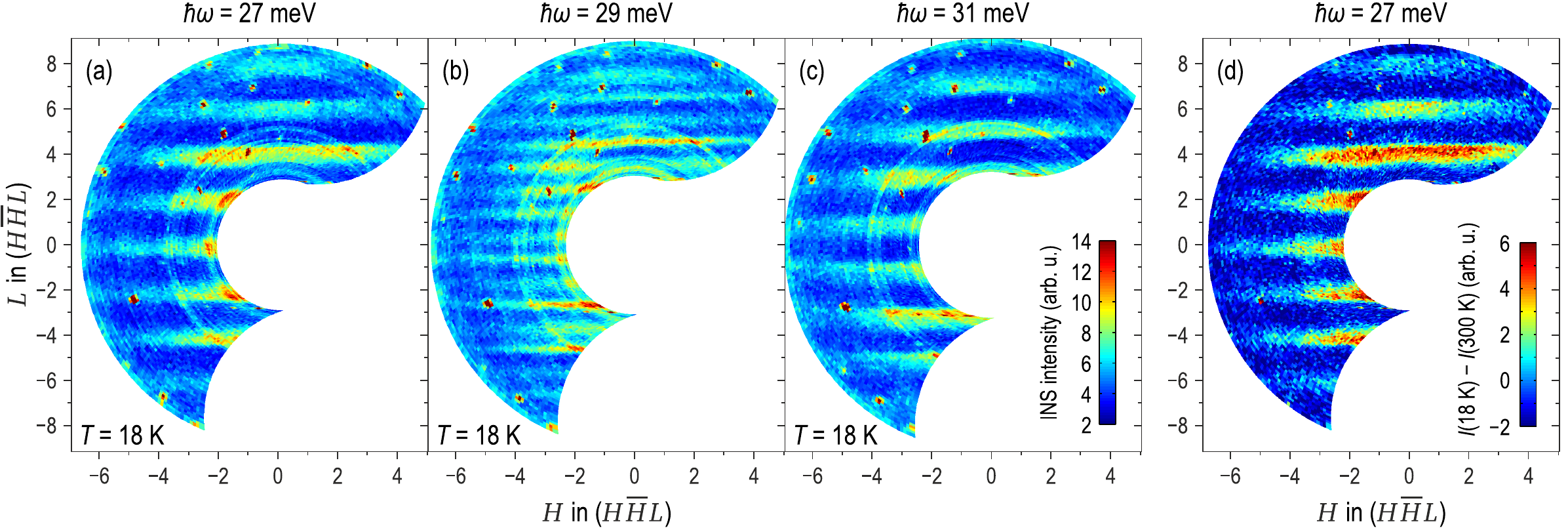}
\caption{(a)--(c)~Constant-energy INS maps along the $(H\kern.5pt\overline{H}\kern.5ptL)$ plane in reciprocal space, measured in the AFM state ($T=18$~K) at three different energies indicated above each panel. The tail of the direct neutron beam and the Al powder lines have been subtracted from the data. The sharp spurious intensity spots originate from accidental Bragg intensity scattered incoherently by the analyzer and should be neglected. (d)~Intensity difference between $T = 18$\,K and $T=300$\,K data at $\hslash\omega=27$~meV, shown to demonstrate the clean background-free magnetic signal and its form-factor decay with increasing $|\mathbf{Q}|$.}
\label{Fig:Qmaps}\vspace{-0pt}
\end{figure*}

While Ca$_{3}$Co$_{2}$O$_{6}$ has been extensively studied using various experimental probes, the small size of the available single crystals has so far precluded measurements of magnetic excitations in this compound by inelastic neutron scattering (INS), which could enable a quantitative estimate of the exchange couplings and single-ion anisotropy. In this Letter, we use the INS technique to directly probe the magnetic excitations in Ca$_{3}$Co$_{2}$O$_{6}$. For these measurements, we meticulously coaligned $\sim$\,400\,mg of small needle-shaped crystals into a mosaic on an Al plate. The $\mathbf{c}$-axis mosaicity of the aligned crystal assembly was $\lesssim$\,3.0$^\circ$. The AFM transition at $\sim$\,24\,K has been confirmed by magnetic susceptibility \cite{Supplemental} and neutron diffraction, in agreement with earlier results \cite{MaignanMichel00, HardyLambert03}. We performed the INS measurements at the thermal-neutron triple-axis spectrometer IN8 (ILL, Grenoble, France), which was operated in the \textit{Flatcone} multi-analyzer configuration \cite{FlatconeNote}. The sample was mounted with its $[1\kern.5pt\overline{1}\kern.5pt0]$ and $[0\kern.5pt0\kern.5pt1]$ directions in the horizontal scattering plane.

In Fig.\,\ref{Fig:Qmaps}\,(a\,--\,c), we show constant-energy maps of the inelastic intensity distribution along the $(H\kern.5pt\overline{H}\kern.5ptL)$ plane in the reciprocal space at three energy transfer values, $\hslash\omega=27$, 29 and 31~meV, in the AFM phase ($T\, =\,18$ K). The wave-vector coordinates are given in reciprocal lattice units (r.l.u.), defined as 1\,$\text{r.l.u.} = 4\piup/\sqrt{3}a$ for the $(H\kern.5pt\overline{H}\kern.5pt0)$ direction and 1\,$\text{r.l.u.} = 2\piup/c$ for the $(0\kern.5pt0\kern.5ptL)$ direction, where $a=9.079$\,\AA\ and $c=10.38$\,\AA\ are the lattice parameters in the hexagonal setting \cite{FjellwagGulbrandsen96}. The tail of the direct neutron beam, centered at $|\mathbf{Q}|=0$, and the Al powder lines originating from the sample holder have been subtracted from the data. At $\hslash\omega$\,=\,27, 29 and 31~meV [Fig.\,\ref{Fig:Qmaps}\,(a\,--\,c)], 1D intensity streaks orthogonal to the chain direction are observed at even, half-integer, and odd $L$ values, respectively, corresponding to the bottom, middle, and top of a dispersive magnon branch. The observed decrease in their intensity towards higher momentum transfer, $|\mathbf{Q}|$, is consistent with the Co$^{3+}$ magnetic form factor, thereby confirming the magnetic origin of these excitations. To illustrate this more clearly, in Fig.\,\ref{Fig:Qmaps}\,(d) we additionally show an intensity difference map [$I(18\text{K})-I(300\text{K})$] for $\hslash\omega = 27$\,meV, where the intensity measured at $T=300$\,K (which is featureless in $\mathbf{Q}$) has been used as a background. Here the intensity streaks are more pronounced in the absence of the background contamination, showing an anisotropic form-factor decay of intensity towards higher $|\mathbf{Q}|$. The absence of any notable dispersion along the $(H\kern.5pt\overline{H}\kern.5pt0)$ direction in Fig.\,\ref{Fig:Qmaps} could originate from very small interchain interactions and/or their effective cancelation due to the geometric frustration. Assuming that a weak dispersion in the $(H\kern.5pt\overline{H}\kern.5pt0)$ direction with a bandwidth smaller than half of the experimental reso\-lution would be unobservable, we can put an upper bound of $\sim$\,0.5\,meV on the dispersion perpendicular to the chains.

\begin{figure}[t]\vspace{-3pt}
\includegraphics[width=\columnwidth]{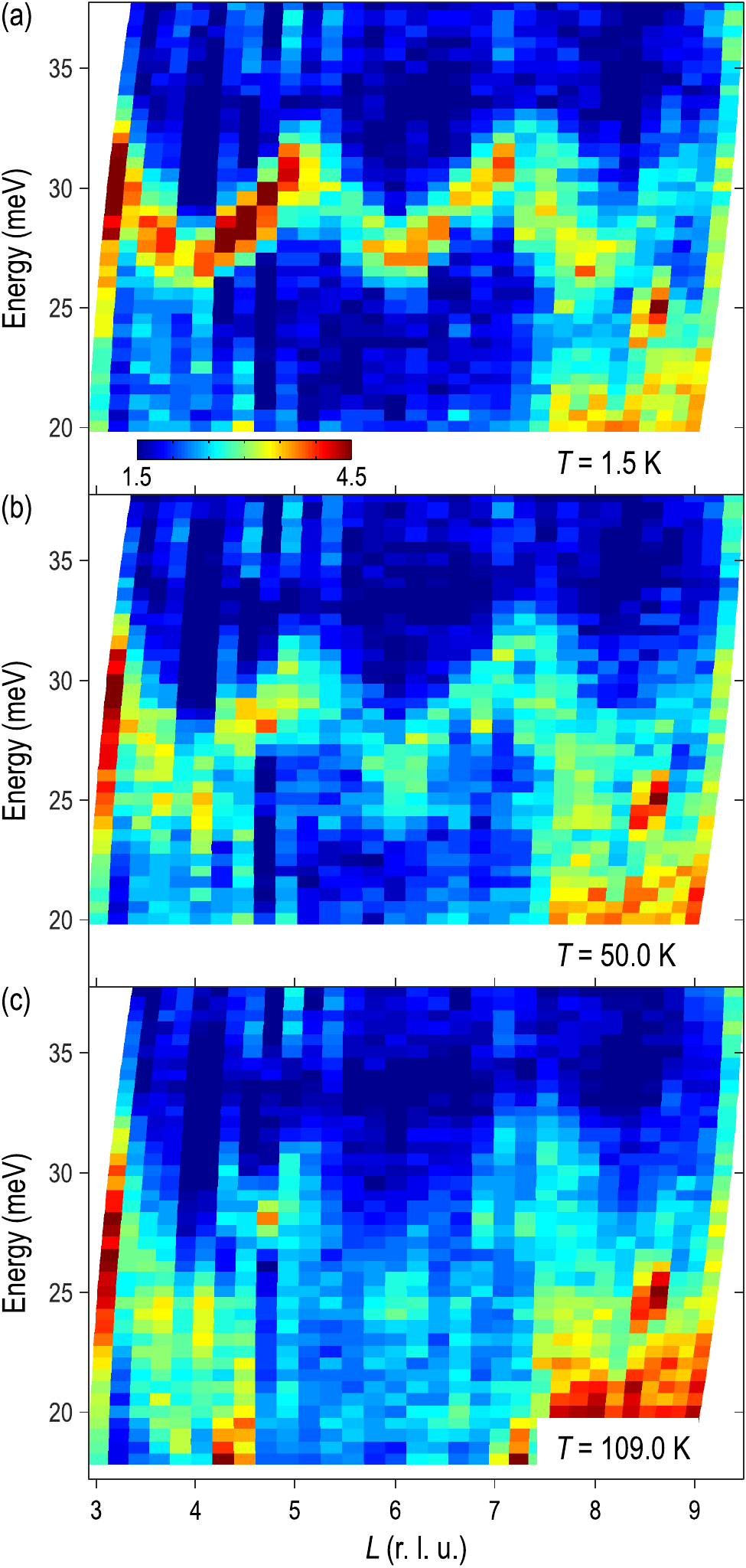}
\caption{Intensity color maps of the observed inelastic neutron spectrum at (a)~1.5\,K, (b)~50\,K, and (c)~109~K. The data are projected onto the $(0\,0\,L)$ direction of the momentum. The $T$-independent contributions from the tail of the direct beam at small $|\mathbf{Q}|$ and Al powder lines from the sample holder around $L=4.5$ have been subtracted from the data.}
\label{Fig:Escan}
\end{figure}

To study the magnon dispersion along the chains, we performed energy scans around $\mathbf{Q} =(0\kern.5pt0\kern.5ptL)$ at various temperatures, both above and below $T_\text{N}$. In the \textit{Flatcone} configuration, every energy scan follows a curved 2D surface in the momentum-energy space. However, since the magnetic signal does not depend on the momentum component orthogonal to the chains, in Fig.\,\ref{Fig:Escan} we present the data in the two-dimensional $(L,\,\hslash\omega)$ projection, neglecting the $H$ component of the momentum without any loss of information. Three representative spectra at $T=1.5$, 50 and 109\,K are shown in Fig.\,\ref{Fig:Escan}\,(a\,--\,c). Additional temperature frames are also presented in the supplemental material (SM)~\cite{Supplemental}. Our main result illustrated by these figures is the observation of a dispersive narrow magnon band with a large spin gap, which ``melts'' gradually upon warming. The average magnon energy, $E_\text{0}$, and its $T$-dependence are consistent with the location of an inelastic peak in earlier INS data measured on a powder sample \cite{CroweAdroja05}. The dispersion reaches its minimum at the FM zone center (even $L$), with a spin gap of $\sim$\,27\,meV at 1.5\,K, and its maximum at the zone boundary (odd $L$) near $\sim$\,30.5\,meV, thereby explaining the location of the intensity streaks in Fig.\,\ref{Fig:Qmaps}. The bandwidth of the magnon branch along the $\mathbf{c}^\ast$-direction therefore amounts to $\sim$\,3.5\,meV. The large ratio of the spin gap to the bandwidth ($\sim$\,8) and the absence of dispersion perpendicular to the chains indicate the highly 1D nature of this material in spin space and in real space, respectively. Above $T_\text{N}$, the gap in the magnetic excitation spectrum starts to fill gradually, yet the magnon band persists at least up to 150\,K [see Fig.\,S1\,(b,\,c) in the SM].

To analyze the magnon dispersion along the chains, we used an effective spin Hamiltonian that includes both the exchange and single-ion anisotropy terms:\vspace{2pt}
\begin{equation}
H = - \sum_i\,[J_{z}\,S_{i}^\emph{z}\,S_{i+1}^\emph{z}\,+\,J_{xy}\,(S_{i}^\emph{x}S_{i+1}^\emph{x} + \,S_{i}^\emph{y}S_{i+1}^\emph{y})\, + D \,(S_{i}^\emph{z})^2]\,.
\label{Eq:Hamiltonian}\vspace{-3pt}
\end{equation}
Here \emph{i} is the site index for the Co$^{3+}_\text{TP}$ ions in the chain, and the \emph{z}-axis is directed along the chains. The first and second terms in Eq.\,\ref{Eq:Hamiltonian} describe the coupling between longitudinal ($J_{z}$) and transverse ($J_{xy}$) components of neighboring Co$^{3+}_\text{TP}$ spins, respectively, and $D$ is the single-ion anisotropy for the Co$^{3+}_\text{TP}$ ions. The AFM interchain interactions were excluded from Eq.\,\ref{Eq:Hamiltonian} for simplicity, as they are very weak and thus do not affect the dispersion along the chains significantly.

Due to the high value of the Co$^{3+}_\text{TP}$ spins, magnetic excitations with different spin quantum numbers $(|{\scriptstyle\Delta}S| \leq 2S)$ are possible in the system. For instance, the simplest excited state of an individual FM chain (neglecting the interchain interactions) is realized when a single spin flips, creating two domain walls, which corresponds to a magnetic excitation with $|{\scriptstyle\Delta}S| = 4$ whose energy is independent of $D$. Although all of these excitations may contribute to thermodynamic properties, the selection rule dictated by the neutron spin restricts the excitations observable by INS to only those with ${\scriptstyle\Delta}S =\pm1$. Within the framework of the conventional linear spin-wave theory, their dispersion is given by
\begin{equation}
\hslash \omega_L = 2S [(D + J_{z}) - J_{xy} \cos({\piup}L) ]\,.
\label{Eq:Dispersion}
\end{equation}

\begin{figure}[t]\vspace{-7pt}
\includegraphics[width=\columnwidth]{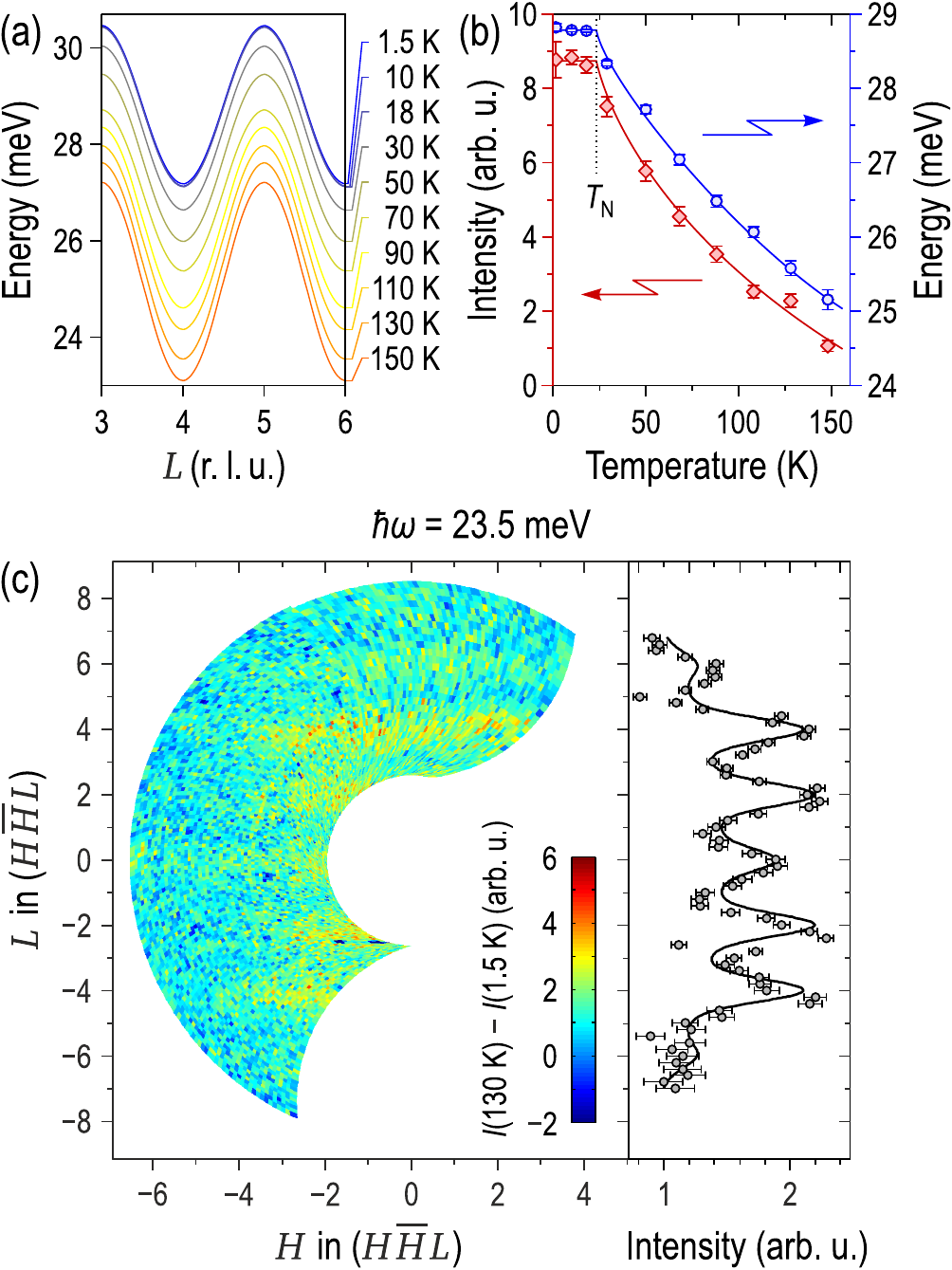}
\caption{ (a)~Fitted magnon dispersion curves along the chains at different temperatures, as indicated in the legend. Background substraction at every temperature was carried out in an iterative manner~\cite{Supplemental}. (b)~The overall peak intensity (diamond symbols) and the energy, $E_{0}$, corresponding to the center of the magnon dispersion curves (circles) obtained from the fit shown in panel (a). (c)~Reciprocal space map at $\hslash\omega=23.5$~meV after subtraction of the background measured inside the spin gap at $T=1.5$\,K, shown to demonstrate the transfer of spectral weight to lower energies upon warming. The projection of intensity onto the $L$ axis is shown to the right.}
\label{Fig:Fit}
\end{figure}

By fitting the experimental data to this model, as described in the SM \cite{Supplemental}, we obtained the magnon dispersion for all measured temperatures, shown in Fig.\,\ref{Fig:Fit}\,(a). At $T=1.5$\,K, the following parameter values resulted from our fitting procedure: $D+J_{z}=7.20\pm0.02$\,meV and $J_{xy}=0.42\pm0.01$\,meV (assuming $S=2$). We note that $J_{z}$ and $D$ enter Eq.\,(\ref{Eq:Dispersion}) additively and therefore cannot be extracted separately from the magnon dispersion alone. Fig.\,\ref{Fig:Fit}\,(b) also shows the $T$-dependence of the main fitting parameters: the overall peak intensity and the center of the magnon band, $E_\text{0}$. The observation of well defined magnetic excitations throughout the Brillouin zone up to at least 150\,K suggests that FM correlations along the chains develop prior to the onset of the 3D magnetic LRO at $T \gg T_\text{N}$. The intensity of the signal, the magnon energy and the spin gap remain essentially constant below $T_{\rm N}$, but start decreasing upon warming above $T_\text{N}$, while the bandwidth remains nearly $T\kern-.5pt$-independent. At $T\geq200$\,K, the signal broadens and is reduced in intensity below the statistical noise level in our data.

\begin{figure}[t]
\includegraphics[width=\columnwidth]{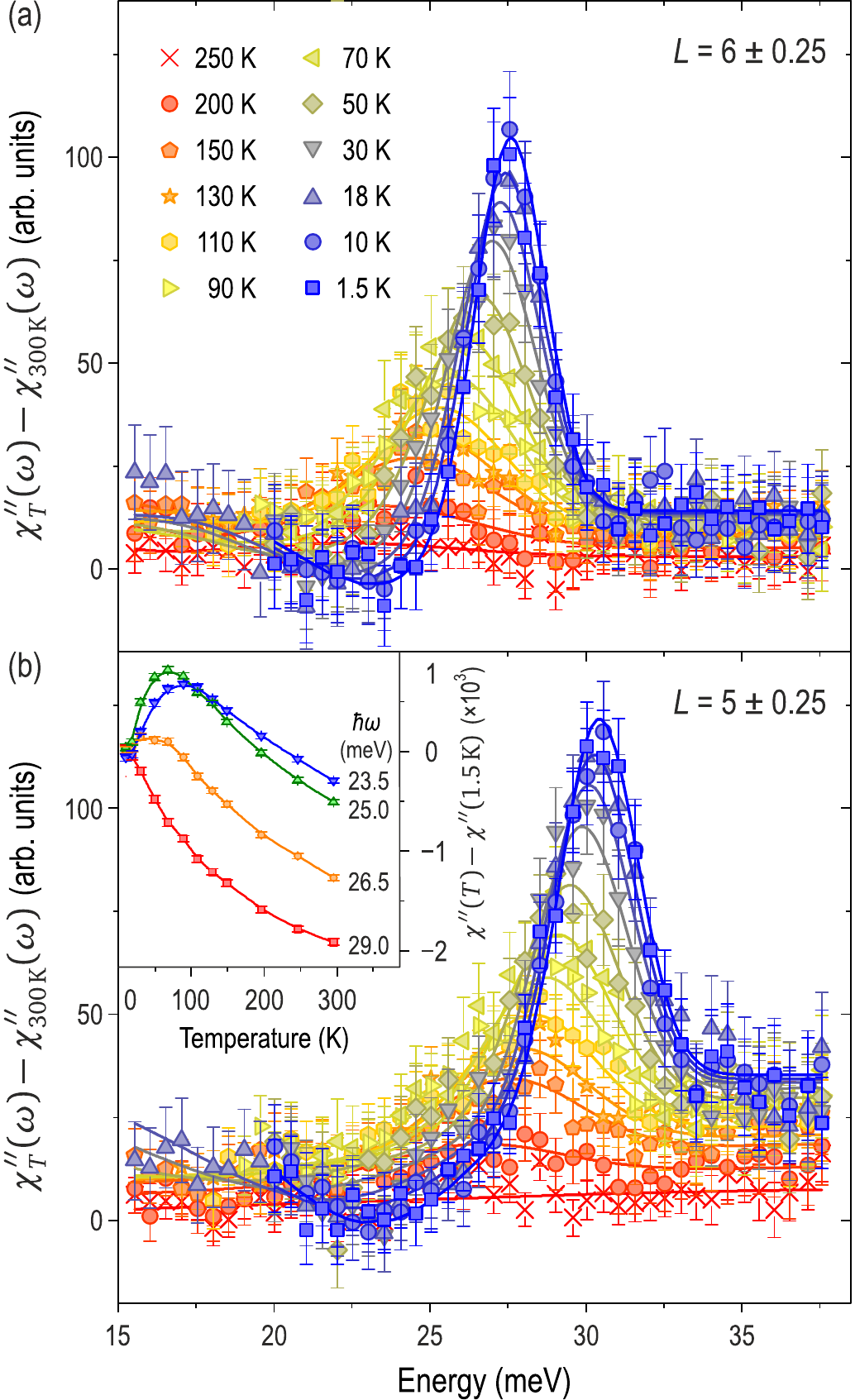}
\caption{Temperature variation of $\chi''_T(\omega)$, integrated over the wave vector range $ \pm 0.25$~r.l.u. around (a)~$L\kern1.2pt$\,=\,$\kern1.2pt5$ and (b)~$L\kern1.2pt$\,=\,$\kern1.2pt6$. The background observed at 300~K was subtracted from the data. Solid lines are empirical fits to the data. The inset shows $T$-dependent changes in the full $\mathbf{Q}$-integrated INS intensity with respect to its low-$T$ values, averaged over $\pm0.5$~meV around $\hslash\omega=23.5$, 25, 26.5, and 29~meV. These data were obtained by $\mathbf{Q}$- and $\omega$-integration from energy scans such as those shown in Fig.\,\ref{Fig:Escan}.}
\label{Fig:Spectweight}
\end{figure}

To demonstrate the softening of the magnon branch upon warming directly, without relying on the model fits, in Fig.\,\ref{Fig:Fit}\,(c) we show the intensity difference, $I(130\,\text{K})-I(1.5\,\text{K})$, taken at $\hslash\omega=23.5$\,meV. At $T=1.5$\,K, this energy lies deep inside the spin gap, as one can see from Fig.\,\ref{Fig:Escan}\,(a), and therefore the base-temperature data serve here as a measure of background to obtain the clean magnetic signal at $T=130$\,K. The observed intensity streaks in Fig.\,\ref{Fig:Fit}\,(c) confirm the shift of the magnetic spectral weight from higher energies down into the spin-gap region, which can be attributed to the reduction of the intrachain FM correlations by thermal excitations.\enlargethispage{1pt}

In Figs.\,\ref{Fig:Spectweight}\,(a,\,b), we examine the spectral-weight redistribution in more detail by following the $T\kern-.5pt$-dependence of the dynamic spin susceptibility, $\chi''_T(\omega)$, momentum-integrated within $\pm 0.25$\,r.l.u. around the bottom ($L\!=\!6$) and top ($L\!=\!5$) of the magnon band, respectively. The high-temperature background observed at 300\,K has been subtracted to emphasize the changes in the magnetic signal with respect to ambient temperature. At 1.5\,K, sharp peaks can be seen near 27.5 and 30.5\,meV for even and odd $L$, respectively, in agreement with the fitted magnon dispersion. These peaks gradually broaden and shift to lower energies upon warming, illustrating the softening of the magnon mode and the reduction of the spin gap. In the inset to Fig.~\ref{Fig:Spectweight}~(b), we also present the change in the $\mathbf{Q}$-integrated $\chi''_T(\omega)$ within several energy windows, averaged over $\pm0.5$~meV around $\hslash\omega=23.5$, 25, 26.5, and 29\,meV, with respect to its low-$T$ values. The intensity at low energies experiences an initial enhancement due to the reduction of the spin gap, while the intensity within the magnon band (29\,meV) is reduced.

We conclude by comparing our results with earlier complementary measurements. A spin gap of $\sim$\,100\,K (8.6\,meV) was previously estimated from {$^{59}$Co} spin-lattice relaxation rates~\cite{Allodi104408}. Intrachain FM correlations above $T_{\rm N}$ have also been reported from the M\"ossbauer effect (in Eu-doped Ca$_3$Co$_2$O$_6$) \cite{PauloseMohapatra08}, zero-field muon-spin relaxation~\cite{Takeshita034712} and Raman-scattering~\cite{Gohil103517} measurements. These correlations are dynamic in nature and do not result in any enhancement of intensity at the nuclear Bragg positions, according to numerous neutron-diffraction studies \cite{KageyamaYoshimura97, Kageyama357, AaslandFjellwag97, AgrestiniFleck11, AgrestiniChapon08}.
Our measured value of the spin gap in Ca$_{3}$Co$_{2}$O$_{6}$ is similar to the one determined recently for the 1D copper-iridium oxide Sr$_3$CuIrO$_6$ using resonant inelastic x-ray scattering \cite{YinLiu13}, where it was attributed to an Ising-like exchange anisotropy, while the band width in the latter compound is approximately 6 times larger than in Ca$_{3}$Co$_{2}$O$_{6}$. It is appropriate to note that the presence of magnetic excitations above $T_\text{N}$ were also reported in other FM spin-chain compounds, such as CsNiF$_{3}$~(Ref.~\citenum{Steiner537}) or CoCl$_{2}$.2D$_{2}$O~(Ref.~\citenum{Kjems5190}). However, the persistence of the magnons up to temperatures as high as $\sim6T_\text{N}$ has not been observed so far in any material, to our knowledge.

The implications of our results can also be considered in the context of a broader class of 1D cobalt chain systems, such as mono-atomic nanowires~\cite{Gambardella301} or single-chain magnets~\cite{Bogani207204, Caneschi771}. As these systems are highly anisotropic in spin space and purely 1D in real space, they cannot exhibit any LRO at nonzero temperature. Still, for Co-nanowires, extremely strong FM correlations persist above the Curie temperature ($T = 0$), and the corresponding relaxation times are long enough to be studied by scanning tunneling spectroscopy and XMCD~\cite{Gambardella301}. For single-chain magnets, the relaxation of the magnetization has been observed in the ac susceptibility~\cite{Bogani207204, Caneschi771}. Yet, these techniques do not yield quantitative information about the microscopic magnetic interactions, which we were able to extract here from neutron-spectroscopy data on a bulk analogue.

In summary, we have observed a dispersing magnetic excitation in Ca$_{3}$Co$_{2}$O$_{6}$, whose quasi-1D character evidences the prevalence of FM intrachain over AFM interchain interactions. The magnon dispersion along the chains, exhibiting an anisotropy gap 8~times larger than the bandwidth, allowed us to give a direct estimate of the anisotropy and exchange parameters in this system, which proves the strongly anisotropic Ising-type character of its magnetic moments.

We thank D.~T.~Adroja, C.~D.~Batista, G.~Khaliullin, C.~Mazzoli, and A.~Zheludev for stimulating discussions. P.~Y.~P. acknowledges financial support by DFG within the Graduiertenkolleg GRK\,1621. H.~J. was supported by the Max Planck POSTECH Center for Complex Phase Materials with KR2011-0031558.

\onecolumngrid\clearpage

\pagestyle{plain}

\renewcommand\thefigure{S\arabic{figure}}
\renewcommand\thetable{S\arabic{table}}
\renewcommand\theequation{S\arabic{equation}}
\renewcommand\bibsection{\section*{\sffamily\bfseries\footnotesize Supplementary References\vspace{-6pt}\hfill~}}

\citestyle{supplement}

\setcounter{page}{1}\setcounter{figure}{0}\setcounter{table}{0}\setcounter{equation}{0}

\pagestyle{plain}
\makeatletter
\renewcommand{\@oddfoot}{\hfill\bf\scriptsize\textsf{S\thepage}}
\renewcommand{\@evenfoot}{\hfill\bf\scriptsize\textsf{S\thepage}}
\renewcommand{\@oddhead}{\Large\textsf{A.~Jain \textit{et~al.}}\hfill\textsf{Supplemental Material}}
\renewcommand{\@evenhead}{\Large\textsf{A.~Jain \textit{et~al.}}\hfill\textsf{Supplemental Material}}
\makeatother

%\makeatletter\immediate\write\@auxout{\string\bibstyle{my-apsrev}}\makeatother

\normalsize

\begin{center}{\vspace*{0.1pt}\Large{Supplemental Material to the Letter\smallskip\\\sl\textbf{``\hspace{1pt}One-Dimensional Dispersive Magnon Excitation\\in the Frustrated Spin-2 Chain System Ca$_3$Co$_2$O$_6$\!''}\smallskip\\by A.~Jain~\textit{et al.}}}\end{center}

\begin{figure}[b!]
\begin{minipage}[b]{0.9\textwidth}
\noindent\animategraphics[controls, loop, width=\textwidth, timeline=timeline.txt]{2}{binder}{1}{13}
\end{minipage}\hfill
\begin{minipage}[b]{0.08\textwidth}
\raisebox{18.33em}{
\begin{tabular}{l}
\includegraphics{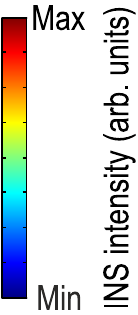}\\
\vspace{10.82em}\\
\includegraphics{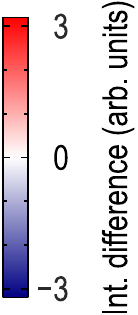}
\end{tabular}}
\end{minipage}
\bigskip
\caption{Animations illustrating the temperature dependence of the magnon dispersion in Ca$_3$Co$_2$O$_6$ along the ferromagnetic chain direction. (a)~Raw INS data after subtraction of Al powder lines. (b)~Background-subtracted data. Here, in every temperature frame we have subtracted the background contribution (data minus the fitted model) evaluated from the next lower temperature. (c)~Temperature subtraction of the neutron scattering intensity, $I(T)-I(\text{300\,K})$, that illustrates the transfer of magnetic spectral weight towards higher energies upon cooling. (d)~Analogous subtraction for $I(T)-I(\text{18\,K})$, illustrating the transfer of spectral weight to lower energies upon warming.\vspace{-2.0em}}
\label{Fig:Animation}
\end{figure}

{\twocolumngrid\parfillskip=0pt\relax
Here in Fig.\,\ref{Fig:Animation} we present an animation showing the full set of INS data on Ca$_3$Co$_2$O$_6$ as a function of temperature. The controls at the bottom of the figure allow to modify the frame rate, pause the animation, or browse through individual temperature frames. Panel (a) shows the raw INS data from energy scans after subtraction of Al powder lines, such as those presented in Fig.\,2 in the paper. Panel (b) shows background-subtracted data. As the estimate for background in every frame, we used data from the next lower temperature, from which we subtracted the fitted model of the magnon dispersion. In the first iteration, we fitted the lowest-temperature data at $T=1.5$\,K with a two-dimensional model function that included the energy of the magnon, the amplitude of the dispersion, the peak width in energy (which was periodically modulated to account for the resolution effects), and the overall peak intensity as free fitting parameters. The model was multiplied by the Co$^{3+}$ magnetic form factor to account for the decay of the signal with increasing $|\mathbf{Q}|$. After that, the result of the fit was subtracted from the data to get an estimate of background for the next temperature frame ($T=10$\,K). This background was subtracted from the $T=10$\,K data, and the resulting dataset, shown in panel (b), was fitted to the same model to obtain the corresponding dispersion.
\onecolumngrid\vfill}

\clearpage

\vspace*{0pt}
\twocolumngrid

\noindent This model was then subtracted from the $T=10$\,K data to obtain the background for the next temperature frame ($T=18$\,K), and so on. Following this procedure, the dispersion at high temperatures, where the signal is masked by the uneven background, could be determined more reliably and up to higher temperatures than would be otherwise possible by fitting the raw data. The resulting dispersion curves and temperature-dependent parameters are plotted in Fig.\,4 of the main paper.

Animations in Fig.\,\ref{Fig:Animation}\,(c,\,d) show temperature subtractions of the raw data as an alternative way of eliminating the effects of the inhomogeneous background. In panel (c), we have subtracted the highest-temperature dataset ($T=300$\,K) from all frames to illustrate the formation of the magnon upon cooling, as the spectral weight from the spin-gap region is gradually transferred to higher energies, where the dispersive magnetic mode starts forming below 200\,K. In panel (d), we have similarly subtracted the 18\,K dataset (immediately below $T_{\rm N}$) from all frames to illustrate the shift of the spectral weight towards lower energies while the mode is destroyed upon warming. In this representation of the data, one can clearly see the formation of a periodic intensity modulation within the spin-gap region (between 15 and 25 meV) at temperatures above $T_{\rm N}$ due to the gradual softening of the mode energy. The same structure can be also seen in the subtraction of constant-energy maps at 23\,meV that is shown in Fig.\,3\,(c) in the paper, as well as directly in the results of the fits presented in Figs.\,3\,(a) and (b).

\begin{figure}[t!]
\includegraphics[width=\columnwidth]{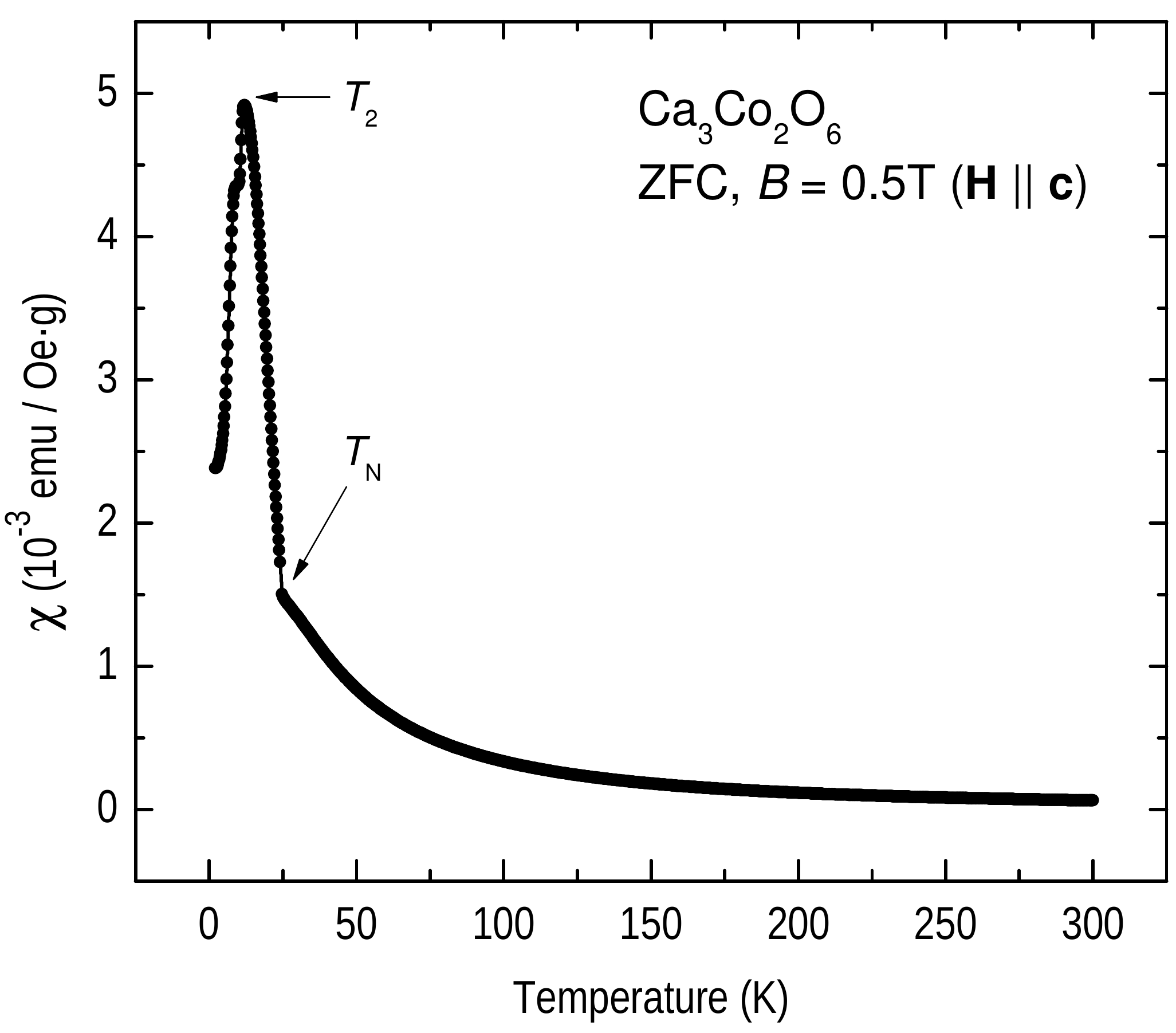}
\caption{Temperature dependence of the static magnetic susceptibility, measured after zero-field cooling in a constant magnetic field of $0.5$\,T applied parallel to the chain direction.}
\label{Fig:Magnetization}\vspace{7pt}
\end{figure}

In Fig.\,\ref{Fig:Magnetization}, we additionally show the results of a magnetic susceptibility measurement done by means of a SQUID magnetometer on one of the Ca$_3$Co$_2$O$_6$ single crystals from the same batch as those used for the neutron mosaic sample. The field was applied parallel to the sample's $\mathbf{c}$-axis after zero-field cooling (ZFC). The anomalies observed at $T_{\rm N}$ and at $T_2\approx12$\,K agree with those reported previously on similar samples \cite{MaignanMichel00, HardyLambert03, AgrestiniFleck11}.

\onecolumngrid

\end{document}